\documentclass[fleqn,usenatbib]{mnras}

\usepackage[T1]{fontenc}

\DeclareRobustCommand{\VAN}[3]{#2}
\let\VANthebibliography\thebibliography
\def\thebibliography{\DeclareRobustCommand{\VAN}[3]{##3}\VANthebibliography}

\usepackage{graphicx}	
\usepackage{amsmath}	
\usepackage{amssymb}	

\usepackage{newtxtext,newtxmath}

\newcommand{\dpmin}{\dot\Pi_\mathrm{min}}
\newcommand{\etab}{\eta_\mathrm{B}}
\newcommand{\itp}{i_\textrm{TP}}
\newcommand{\Mzams}{M_\mathrm{ZAMS}}
\newcommand{\tev}{t_\textrm{ev}}

\title[Hydrodynamic modelling of T UMi]{Hydrodynamic modelling of pulsation period decrease in the Mira--type variable T UMi}

\author[Yu. A. Fadeyev]{
Yuri A. Fadeyev,$^{1}$\thanks{E-mail: fadeyev@inasan.ru}
\\
$^{1}$ Institute of Astronomy, Russian Academy of Sciences,  119017, Pyatnitskaya str., 48, Moscow, Russia
}

\date{Accepted XXX. Received YYY; in original form ZZZ}

\pubyear{2022}

\begin{document}
\label{firstpage}
\pagerange{\pageref{firstpage}--\pageref{lastpage}}
\maketitle

\begin{abstract}
Pulsation period decrease during the initial stage of the thermal pulse in the helium--burning
shell of the Mira--type variable T UMi is investigated with numerical methods of stellar evolution
and radiation hydrodynamics.
To this end, a grid of evolutionary tracks was calculated for stars with masses on the main
sequence $1M_{\sun}\le \Mzams\le 2.2M_{\sun}$ and metallicity $Z=0.01$.
Selected models of AGB evolutionary sequences were used for determination of the initial
conditions and the time--dependent inner boundary conditions for the equations of hydrodynamics
describing evolutionary changes in the radially pulsating star.
The onset of period decrease during the initial stage of the thermal pulse is shown to nearly
coincide with the peak helium--burning luminosity.
The most rapid decrease of the period occurs during the first three decades.
The pulsation period decreases due to both contraction of the star and mode switching from
the fundamental mode to the first overtone.
The time--scale of mode switching is of the order of a few dozen pulsation cycles.
The present--day model of the Mira--type variable T UMi is the first--overtone pulsator with
small-amplitude semi--regular oscillations.
Theoretical estimates of the pulsation period at the onset of period decrease and the rate
of period change three decades later are shown to agree with available observational data on T UMi
for AGB stars with masses $ 1.04M_{\sun}\le M\le 1.48M_{\sun}$.
\end{abstract}




\section{Introduction}

Variability of T UMi was discovered at the beginning of the twentieth century by
\cite{pf1906} and for a long time this star was known as
a long--period Mira--type variable with fairly regular light variations and the period
$\Pi\approx 300$ d \citep{s2017}.
This star has attracted the attention after reports by \cite{gs1995} and \cite{mf1995} on
significant decrease of the pulsation period which has begun in 1970s
when its period of light variations was $\Pi\approx 315$ d.
The authors of these papers supposed that observed period decrease is due to the star
contraction during the initial stage of the thermal pulse in the helium burning shell
\citep{wz1981}.
Observational estimates of the period change rate obtained from parabolic fits of
the $O-C$ diagram spanning two decades are $\dot\Pi\approx -3.5$~d/yr \citep{s2002}
and $\dot\Pi\approx -3.8$~d/yr \citep{skb2003}.
In 2001 the period of T UMi was $\Pi\approx 230$ d and afterwards the fairly periodic
oscillations changed to semi--regular light variations with possible switch from
the fundamental mode to the first overtone \citep{u2011}.
In more detail observations of T UMi are duscussed by \cite{mjk2019}.

The date of the onset of pulsation period decrease in T UMi is known with accuracy about
10 yr so that this star is of great interest because of an opportunity to obtain constraints
on the mass of the AGB star.
However, only two theoretical studies devoted to evolution and pulsations of T UMi
have been done so far.
\cite{f2018} computed the grid of non--linear pulsation models of red giants with initial
conditions obtained from evolutionary sequences of AGB--stars with masses on the main
sequence $1M_{\sun}\le \Mzams\le 2M_{\sun}$.
He found that the mass of T UMi is $M=0.93M_{\sun}$.
\cite{mjk2019} undertook a linear analysis of stellar pulsations in red giants also based on
stellar evolution calculations and their estimate of the mass of T UMi is $M=1.66M_{\sun}$.

Though the authors of both works used the \cite{b1995} mass loss rate formula,
\cite{f2018} carried out evolutionary calculations with parameter $\etab = 0.05$ whereas
\cite{mjk2019} used the value $\etab=0.1$.
Below we show that evolutionary calculations of the AGB stage with mass loss parameter
in the range $0.02\le \etab\le 0.1$ allow us to obtain hydrodynamic models reproducing
period decrease observed during several decades in T UMi notwithstanding the fact that
their stellar masses range from $1.04M_{\sun}$ to $1.48M_{\sun}$.
Therefore, the existing observational data on T UMi are insufficient for accurate
determination of the stellar mass.
One should also bear in mind that the mass loss rates of AGB stars remain highly uncertain
\citep{w2000,ho2018}.

Another cause of uncertainties in estimates of the stellar mass obtained by \cite{f2018} and
\cite{mjk2019} is due to the assumption that the initial model of the stellar envelope
used for computations of stellar pulsations is assumed to be in thermal equilibrium.
As we show below, this assumption appropriate in studies of Cepheids or RR Lyr stars becomes
wrong in the case of significant deviation from thermal equilibrium in the contracting envelope
of the AGB star undergoing a thermal pulse in the helium--burning shell.

The goal of the present study is to diminish the role of uncertainties mentioned above.
First, the evolutionary sequences of AGB stars are calculated with different values of the
mass loss parameter $\etab$ in order to evaluate dependence of the theoretical estimate
of the stellar mass on the mass loss rate.
Second, the equations of radiation hydrodynamics are solved with time--dependent inner
boundary conditions obtained from evolutionary calculations so that the hydrodynamic model
is fully consistent with evolutionary changes in the stellar envelope.
This method is appropriate for calculating the non--linear pulsations of stars with significant
deviation from thermal equilibrion and has earlier been used for explanation of decaying
oscillations in the type--II Cepheid RU Cam \citep{f2021}.
Hydrodynamic models of TU UMi presented below reproduce evolutionary changes in stellar pulsations
for the time interval $\loa 120$ yr after the onset of period decrease.

\section{Evolutionary sequences of AGB stars}
\label{sec:evolseq}

Initial conditions for solution of the equations of hydrodynamics were determined
on the basis of evolutionary computations for stars with masses on the main sequence
$1M_{\sun}\le \Mzams\le 2.2M_{\sun}$.
The initial fractional abundances of helium and heavier elements were assumed to be
$Y=0.28$ and $Z=0.01$, respectively.
Stellar evolution was computed with the program MESA version 15140 \citep{pss2019}.
Convection was treated according to the standard mixing--length theory \citep{bv1958} with
mixing length to pressure height ratio $\alpha = 1.8$.
The extended convective mixing was treated according to the exponential diffusive overshoot method
\citep{h2000} with parameter $f_\mathrm{ov}=0.014$.
Energy generation rates and nucleosynthesis were computed using the nuclear reaction network
consisting of 26 isotopes from hydrogen ${}^1\textrm{H}$ to magnesium ${^{24}}\textrm{Mg}$.
The rates of 81 reactions were calculated with the database REACLIB \citep{c2010}.

Mass loss rates at evolutionary stages preceding AGB were evaluated by the \citep{r1975} formula
using the parameter $\eta_\mathrm{R}=0.5$.
We assumed that the AGB stage begins when the central helium abundance is $Y_\mathrm{c} \le 10^{4}$.
Calculations of AGB stellar evolution were done for three different values of the mass loss
rate parameter ($\etab = 0.02$, 0.05, 0.1) of the \cite{b1995} formula so that three evolutionary
sequences of the AGB stage were computed for each value of the initial stellar mass $\Mzams$.

\begin{figure}
	\includegraphics[width=\columnwidth]{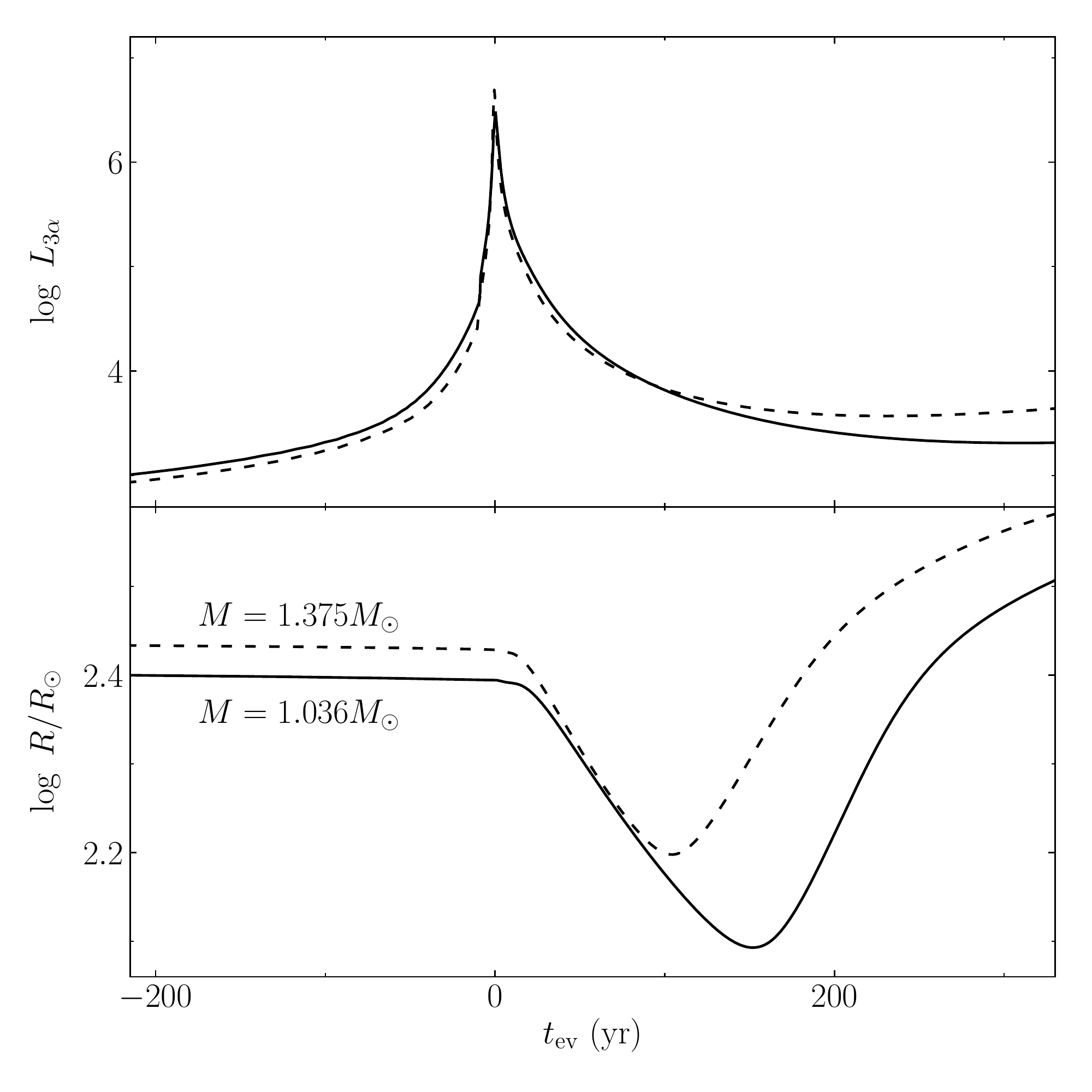}
    \caption{Time dependence of the helium--burning luminosity $L_{3\alpha}$ (top panel) and
    the stellar radius $R$ for evolutionary sequences
    $\Mzams=1.3M_{\sun}$, $\etab=0.02$, $\itp=8$ (solid lines) and
    $\Mzams=1.7M_{\sun}$, $\etab=0.05$, $\itp=10$ (dashed lines).
    The stellar mass $M$ is indicated at the plots.}
    \label{fig:fig1}
\end{figure}

In the present study we assumed that the onset of period decrease coincides with the maximum of the
helium--burning luminosity $L_{3\alpha}$.
The applicability of this approximation is illustrated in Fig.~\ref{fig:fig1}, where
the helium--burning luminosity (top panel) and the radius of the evolving star (bottom panel)
are shown as a function of time $\tev$ for evolutionary sequences
$\Mzams=1.3M_{\sun}$, $\etab=0.02$, $\itp=8$ and $\Mzams=1.7M_{\sun}$, $\etab=0.05$, $\itp=10$.
Here $\itp$ is the number of the thermal pulse and the time $\tev$ is set to zero at the maximum
of $L_{3\alpha}$.
As seen in Fig.~\ref{fig:fig1}, the radius commences to decrease within a time interval
$0\la\tev\la 10$ yr.
Because the pulsation period and the radius relate as $\Pi\propto r^{3/2}$ we can conclude
that this time interval is comparable with observational uncertainty of the onset of period
decrease in T UMi.

To measure the deviation from thermal equilibrium in the spherically symmetric stellar envelope
we used the parameter $\delta_\mathrm{L} = |1 - L_N/L_0|$, where $L_0$ and $L_N$ are the
luminosities at the innermost and the uppermost mass zones of the stellar envelope model,
respectively.
Envelopes of evolutionary models with age $\tev\le 0$ were found to insignificantly deviate
from thermal equilibrium ($\delta_\mathrm{L} \loa 0.02$) so that the non--linear pulsation
model can be computed with the time--independent inner boundary conditions
\begin{equation}
\label{eq:sibc}
 \dfrac{\partial r_0}{\partial t} = \dfrac{\partial L_0}{\partial t} = 0 .
\end{equation}

To compute the limit--cycle pulsation models, we solved the Cauchy problem for the equations
of radiation hydrodynamics \citep{f2015} and time--dependent turbulent convection \citep{k1986}.
The model of the stellar enevelope selected for calculation of initial conditions was divided
into 600 zones with the innermost zone at the radius $r_0\approx 2\times 10^{-3}R$, where $R$
is the radius of the outer boundary.
The mass intervals of 500 outer zones increase geometrically inwards whereas the mass
intervals of 100 inner zones decrease.
The use of smaller mass intervals at the bottom of the envelope is necessary to obtain
a better approximation in the inner layers of the stellar envelope with rapidly
increasing pressure and temperature gradients.

\begin{figure}
	\includegraphics[width=\columnwidth]{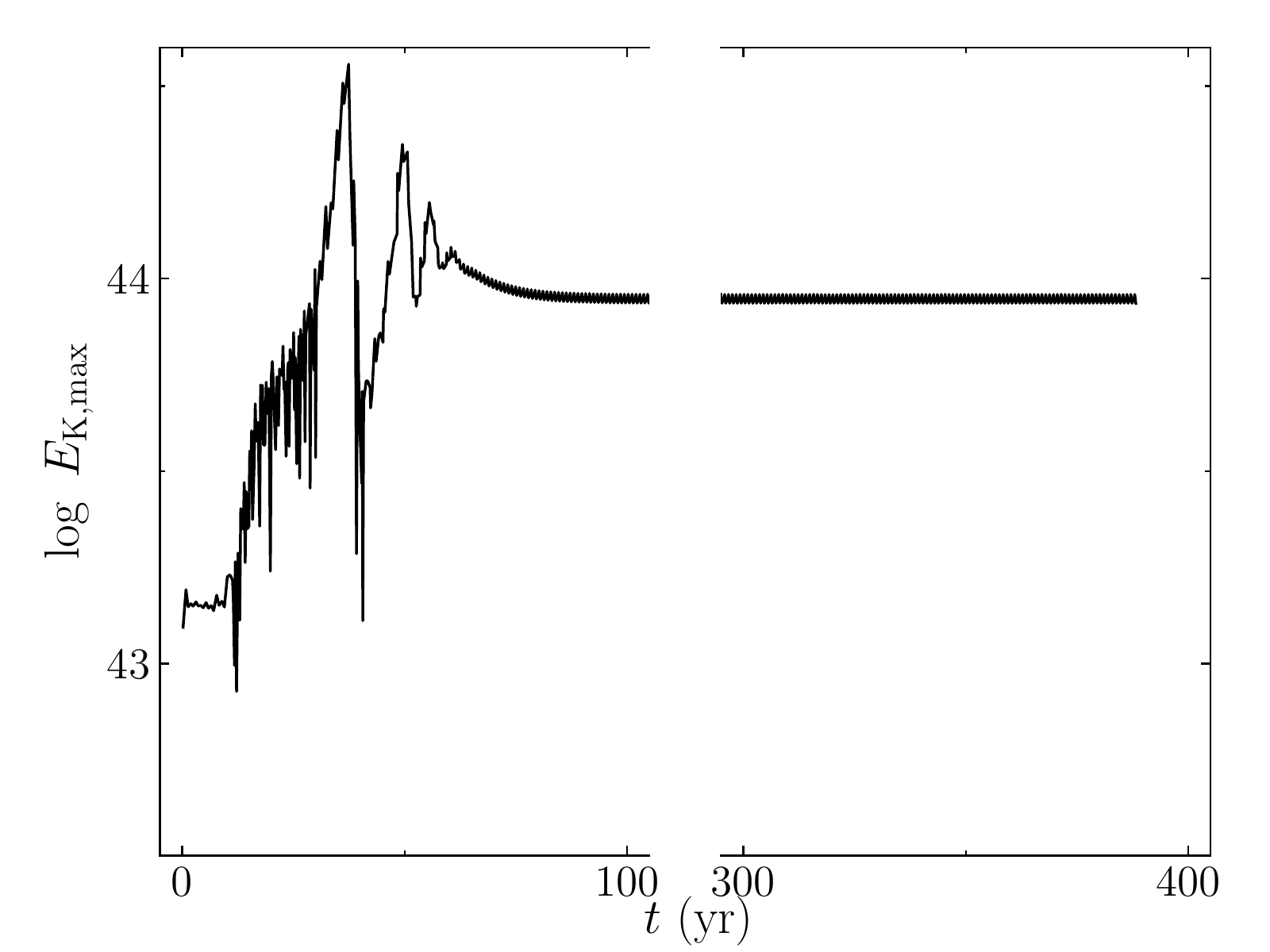}
    \caption{
    The maximum kinetic energy $E_\mathrm{K,max}$ as a function of time $t$ during amplitude growth
    and limit--cycle oscillations of the model $\Mzams=1.8M_{\sun}$, $\etab=0.1$, $\itp=11$.}
    \label{fig:fig2}
\end{figure}

Regular radial pulsations of red giants exist in the form of the standing wave with the open outer
boundary so that the kinetic energy of pulsation motions $E_\mathrm{K}$ reaches the maximum twice
per period.
Fig.~\ref{fig:fig2} shows the maxima of $E_\mathrm{K}$ as a function of time for the model
of the evolutionary sequence $\Mzams=1.8M_{\sun}$, $\etab=0.1$, $\itp=11$
with pulsation period $\Pi^*=312$ d.
Calculations of initial conditions were finished after attainment of the limit--cycle
oscillations.
The pulsation period $\Pi^*$ of the model with age $\tev=0$ was evaluated using the discrete
Fourier transform of the kinetic energy of pulsation motions $E_\mathrm{K}$ at the stage of
limit--cycle oscillations.

\section{Hydrodynamic models}
\label{sec:hydromod}

In this work we selected 26 evolutionary sequences with different values of $\Mzams$, $\etab$ and
$\itp$ for calculation of hydrodynamic models.
Their pulsation periods at the peak helium--burning luminosity range between 262 d and 365 d.
The initial conditions for calculating the hydrodynamic models were determined using the
non--linear limit--cycle pulsation models with the age $-200~\textrm{yr}\loa \tev \loa -100~\textrm{yr}$.
Deviation from thermal equilibrium in the evolutionary model of the stellar envelope is
$\delta_\mathrm{L} < 10^{-3}$ so that calculations of limit--cycle stellar oscillations were performed
in the manner described in the previous section.

\subsection{Inner boundary conditions}
\label{sec:ibcond}

\begin{figure}
	\includegraphics[width=\columnwidth]{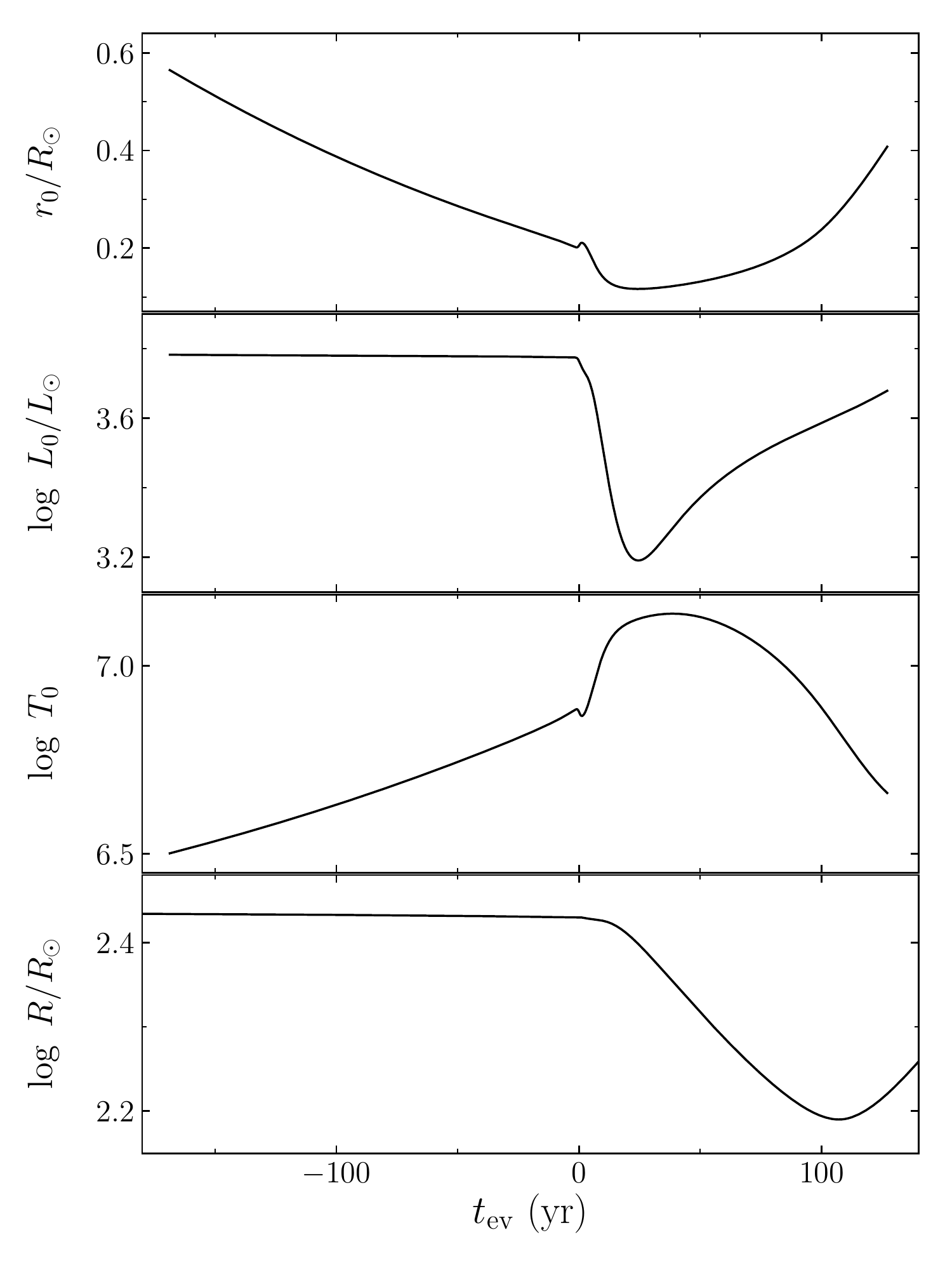}
    \caption{Time dependence of the radius $r_0$, the luminosity $L_0$ and the temperature $T_0$
    at the innermost mass zone of the hydrodynamic model $\Mzams=1.8M_{\sun}$, $\etab=0.1$, $\itp=11$.
    In the bottom panel, the radius $R$ of the evolutionary model as a function of the age $\tev$.}
    \label{fig:fig3}
\end{figure}

Calculation of the hydrodynamic model implies that the time--independent inner boundary conditions
(\ref{eq:sibc}) are replaced by the time--dependent inner boundary conditions $r_0(\tev)$,
$L_0(\tev)$ determined from evolutionary calculations for the Lagrangean coordinate of the innermost
mass zone.
An example of time--dependent inner boundary conditions is displayed in two upper panels of
Fig.~\ref{fig:fig3}.
The plots of $T_0$ and $R$ in two lower panels show time dependences of the temperature at
the innermost mass zone and the radius of the evolutionary model.

In the present study we assumed that no convection occurred at the inner boundary
so that the innermost mass zone should be always deeper than the bottom of the outer
convection zone.
Rapid decrease of the temperature $T_0$ for $\tev > 30$ yr (see Fig.~\ref{fig:fig3})
is accompanied by inward movement of the bottom of the outer convection zone.
Hydrodynamic calculations are carried out only until
the bottom of the convection zone
reaches the inner boundary of the model.
As seen in the lower panel of Fig.~\ref{fig:fig3}, our method of calculations allows us
to consider the whole stage of stellar radius decrease as well as the initial stage of subsequent
expansion of the star.

\subsection{Evolution of stellar pulsations}
\label{sec:evolsp}

The solution of the equations of hydrodynamics with time--dependent inner boundary
conditions is almost fully consistent with calculations of stellar evolution.
The only exception is that the hydrodynamic computations are done for the model with
the constant mass.
However at the end of hydrodynamic computations the difference between masses of
the evolutionary and hydrodynamic models is less than $\sim 0.03\%$ so that
effects of mass loss in hydrodynamic computations can be neglected.

\begin{figure}
	\includegraphics[width=\columnwidth]{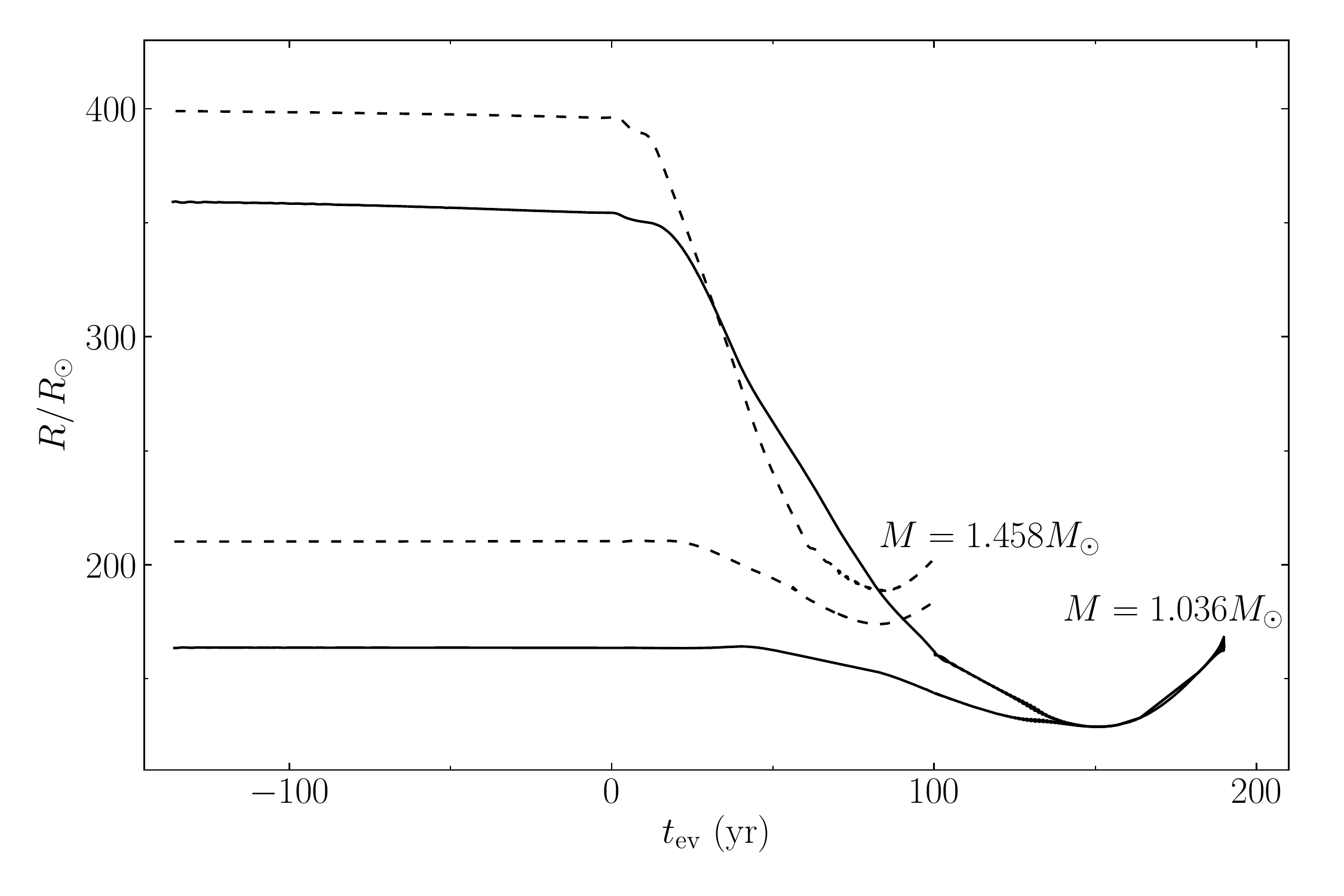}
    \caption{Time dependence of the minimum and maximum radii of hydrodynamic models
    $\Mzams=1.3M_{\sun}$, $\etab=0.02$, $\itp=8$ (solid lines) and
    $\Mzams=1.7M_{\sun}$, $\etab=0.02$, $\itp=14$ (dashed lines).
    Masses of hydrodynamic models are given near the curves.}
    \label{fig:fig4}
\end{figure}

In Fig.~\ref{fig:fig4} we show the minimum and maximum radii of the pulsating star
for hydrodynamic models $\Mzams=1.3M_{\sun}$, $\etab=0.02$, $\itp=8$ and
$\Mzams=1.7M_{\sun}$, $\etab=0.02$, $\itp=14$.
At the age $\tev=0$ the mass of the star, the pulsation period and the relative amplitude at
the upper boundary are $M=1.036M_{\sun}$, $\Pi^*=311$ d, $\delta R/R\approx 0.76$ and
$M=1.458M_{\sun}$, $\Pi^*=330$ d, $\delta R/R\approx 0.61$, respectively.
Decline of the pulsation amplitude during contraction of the star is due to decrease of
the mass of the hydrogen ionization zone accompanying temperature increase in the stellar
envelope.

\begin{figure}
	\includegraphics[width=\columnwidth]{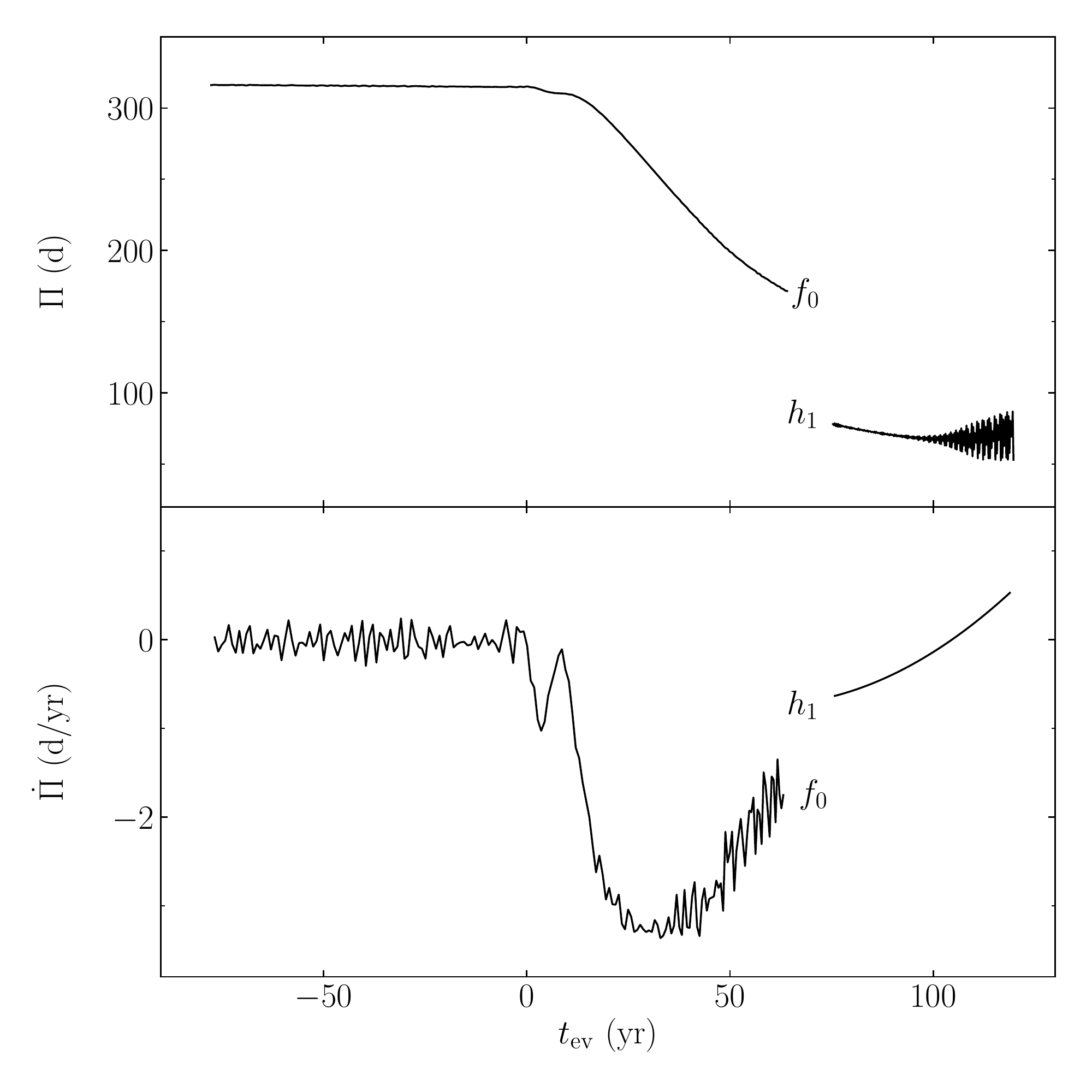}
    \caption{The time dependence of the pulsation period $\Pi$ (upper panel) and
    the rate of period change $\dot\Pi$ (lower panel) for the hydrodynamic
    model $\Mzams=1.8_{\sun}$, $\etab=0.10$, $\itp=11$.
    Plots labelled as $f_0$ and $h_1$ correspond to pulsations in the fundamental
    mode and in the first overtone, respectively.}
\label{fig:fig5}
\end{figure}

The pulsation period of the hydrodynamic model $\Pi(\tev)$ was evaluated for each cycle of
pulsations as a time interval between two adjacent maxima of the radius of the outer boundary.
The temporal dependence of the period for the hydrodynamic model $\Mzams=1.8_{\sun}$, $\etab=0.10$,
$\itp=11$ is shown in the upper panel of Fig.~\ref{fig:fig5}.
The gap in the plot corresponds to switch of oscillations from the fundamental mode ($f_0$)
to the first overtone ($h_1$) when evaluation of the period becomes impossible because of
a beat phenomenon.

In the lower panel of Fig.~\ref{fig:fig5} we show the time dependence of the rate of
period change $\dot\Pi$.
For the time interval with fundamental mode pulsations the period change rate was evaluated
by differentiation of the Lagrange second degree interpolating polynomial and the scatter
of individual estimates of $\dot\Pi$ is due to the fact that radial pulsations are not
perfectly periodic.
At the same time, we did not succeed to calculate in the same manner $\dot\Pi(\tev)$ for
oscillations in the first overtone because of increasing scatter of individual estimates
of the period (see the plot $h_1$ in the upper panel of Fig.~\ref{fig:fig5}).
In the following subsection we will show that this is due to increasing amplitude of
the fundamental mode.
Rough estimates of $\dot\Pi$ shown in the lower panel of Fig.~\ref{fig:fig5} by the
smooth line were obtained by differentiation of the 3rd order algebraic polynomial approximating
the time dependence $\Pi(\tev)$ labelled in the upper panel as $h_1$.

\begin{figure}
	\includegraphics[width=\columnwidth]{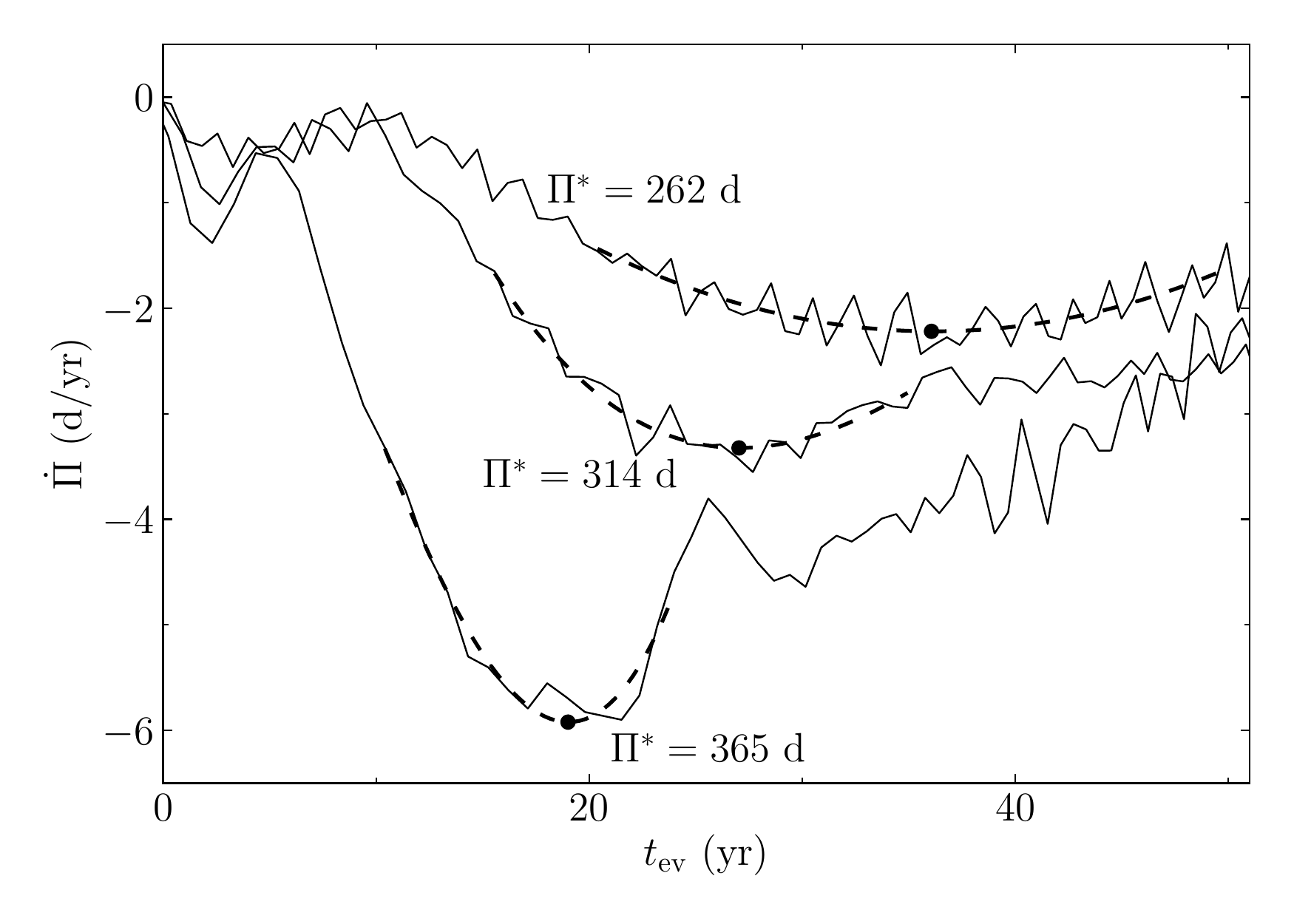}
    \caption{The rate of period change as a function of time for hydrodynamic models with
    period at the maximum helium--burning luminosity $\Pi^*=262$ d, 314 d and 365 d.
    Dashed lines show the polynomial approximation of the minimum of $\dot\Pi$.
    Filled circles represent estimates of the highest rate of period decrease.}
\label{fig:fig6}
\end{figure}

The most conspicuous feature of the plot labelled as $f_0$ in the lower panel of Fig.~\ref{fig:fig5}
is the deep minimum of $\dot\Pi(\tev)$ indicating that the most rapid period decrease
takes place during a few decades after maximum of the helium--burning luminosity.
In Fig.~\ref{fig:fig6} we show the plots of $\dot\Pi(\tev)$ for three hydrodynamic models
with periods at the maximum helium--burning luminosity $\Pi^*=262$ d, 314 d and 365 d.
These models were computed for three different evolutionary sequences: 
$\Mzams=1.6_{\sun}$, $\etab=0.10$, $\itp=7$,
$\Mzams=1.4_{\sun}$, $\etab=0.02$, $\itp=9$ and
$\Mzams=2.2_{\sun}$, $\etab=0.10$, $\itp=19$, respectively.
Estimates of $\dot\Pi$ minima were obtained by polynomial approximation (dashed
lines) and are shown by filled circles.
Among 26 hydrodynamic models with periods $262~\textrm{d}\le\Pi^*\le 365~\textrm{d}$
we found that the highest rate of period decrease $\dpmin$ as well as the time corresponding
to this minimum $\tev(\dpmin)$ depend only on the the value of period $\Pi^*$.

Table~\ref{tab:table} lists properties of eight hydrodynamic models with periods $\Pi^*$ different
from the value 315 d by less than 5\% so that below they are referred as hydrodynamic models of T UMi.
The second and the third columns give the values of the highest rate of period decrease
$\dpmin$ and of the corresponding evolution time $\tev(\dpmin)$.
Columns labelled as $\Pi_0$ and $\Pi_1$ give the periods of the fundamental mode and of
the first overtone evaluated by extrapolation of dependences $f_0$ and $h_1$ with respect to
$\tev$ into the central point of the gap.
It should be noted that the period ratios $\Pi_1/\Pi_0$ given in the sixth column
agree well with estimates obtained from the linear theory \citep{fw1982,xd2007,mjk2019}.
In the seventh column we give the mass of the hydrodynamic model $M$ and in the last three columns
we provide the values of $\Mzams$, $\etab$ and $\itp$.

\begin{table}
	\centering
	\caption{Hydrodynamic models of T UMi.}
	\label{tab:table}
    \begin{tabular}{lccccccccr}
		\hline
        $\Pi^*$ & $\dpmin$ & $\tev(\dpmin)$ & $\Pi_0$ & $\Pi_1$ & $\Pi_1/\Pi_0$ & $M$ & $\Mzams$ & $\etab$ & $\itp$\\
        $\textrm{d}$ & $\textrm{d/yr}$ & $\textrm{yr}$ & $\textrm{d}$ & $\textrm{d}$ & & $M_{\sun}$ & $M_{\sun}$ & &\\
		\hline
        311 &  -3.31 &  31.1 &   142 &    71 & 0.500 & 1.04 &  1.3 &  0.02 &  8 \\ 
        312 &  -3.28 &  29.5 &   164 &    82 & 0.500 & 1.32 &  1.8 &  0.10 & 11 \\ 
        314 &  -3.35 &  26.9 &   150 &    75 & 0.500 & 1.14 &  1.4 &  0.02 &  9 \\ 
        318 &  -3.37 &  29.2 &   174 &    88 & 0.506 & 1.38 &  1.7 &  0.05 & 10 \\ 
        318 &  -3.20 &  31.4 &   151 &    75 & 0.497 & 1.24 &  1.5 &  0.02 & 10 \\ 
        320 &  -3.50 &  30.7 &   169 &    85 & 0.503 & 1.36 &  1.6 &  0.02 & 11 \\ 
        321 &  -3.18 &  31.8 &   170 &    85 & 0.500 & 1.39 &  1.6 &  0.02 & 10 \\ 
        325 &  -3.83 &  28.4 &   177 &    91 & 0.514 & 1.48 &  2.0 &  0.10 & 16 \\ 
		\hline
	\end{tabular}
\end{table}

A cursory glance at Table~\ref{tab:table} reveals that all hydrodynamic models
insignificantly differ from each other in values of $\dpmin$ and $\tev(\dpmin)$.
Moreover, the average values $\langle\dpmin\rangle \approx -3.4$~d/yr and
$\langle\tev(\dpmin)\rangle \approx 30$~yr are in good agreement with observational
estimates $\dot\Pi\approx -3.5$~d/yr \citep{s2002} and $\dot\Pi\approx -3.8$~d/yr \citep{skb2003}
obtained nearly 30 yr after the onset of period decrease.
Thus, the mass of T UMi seems to be in the range $1.04\le M/M_{\sun} \le 1.48$.

\begin{figure}
	\includegraphics[width=\columnwidth]{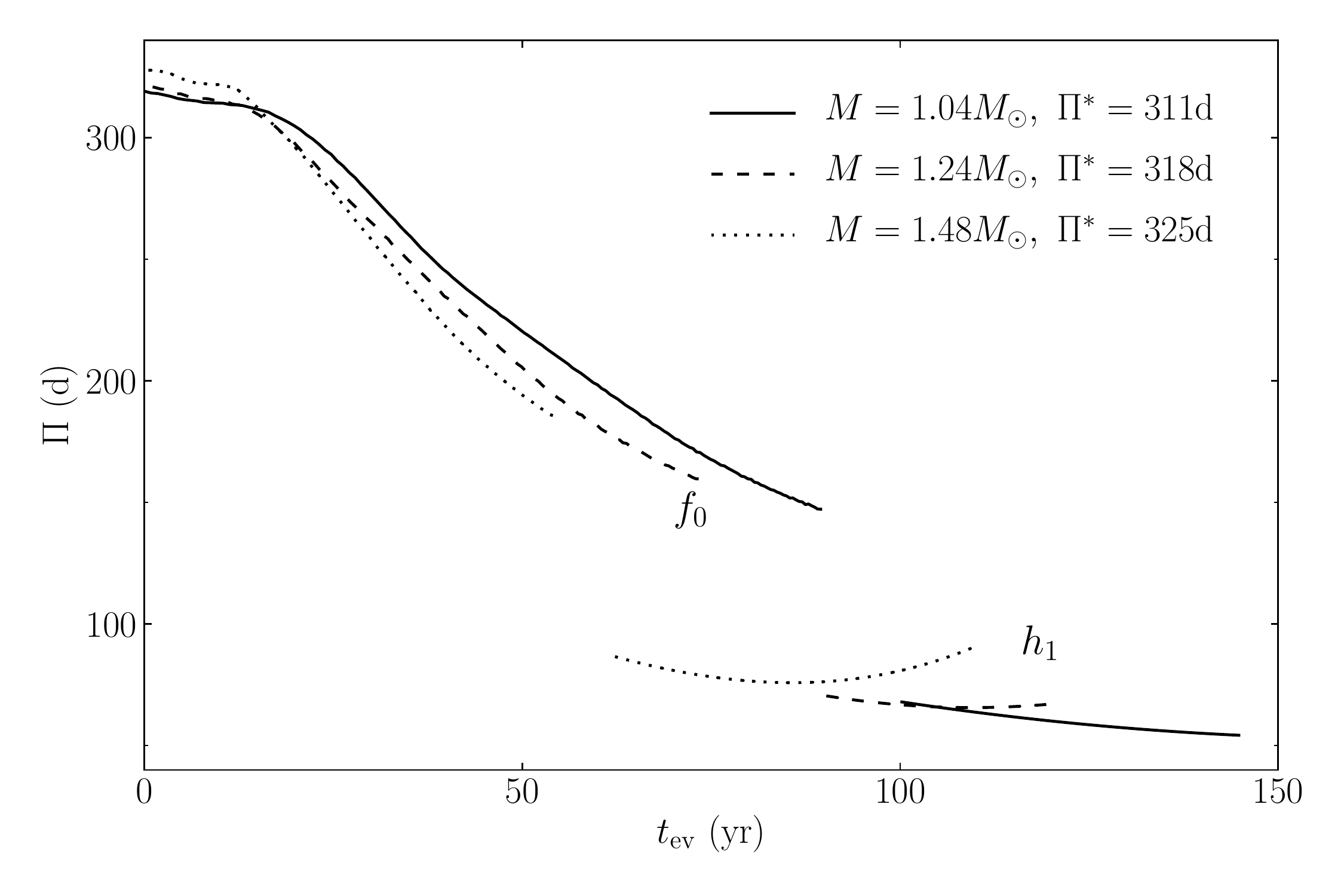}
    \caption{The pulsation period as a function of time for three hydrodynamic models
    listed in Table~\ref{tab:table}.}
\label{fig:fig7}
\end{figure}

More rigorous constraint on the mass can be obtained when cessation of period decrease is
detected observationally.
Dependence of the duration of period decrease on the stellar mass is illustrated in Fig.~\ref{fig:fig7}
where the plots of the period as a function of time $\tev$ are shown for three hydrodynamic models
with different stellar masses.
The minimum of the period is achieved when the star is the first overtone pulsator whereas the
duration of period decrease ranges from $\approx 85$ yr for $M=1.48M_\odot$ to $\goa 150$ yr
for $M=1.04M_\odot$.

In conclusion of this subsection, it is to be noted that as soon as the deviation from thermal equilibrium
in the pulsating envelope becomes appreciably large ($\delta_\mathrm{L} \goa 0.1$)
the use of the time--dependent inner boundary conditions leads to results that significantly differ
from those obtained with (\ref{eq:sibc}).
For example, deviation from thermal equilibrium in the envelope of the model of the evolutionary sequence
$\Mzams=1.8M_{\sun}$, $\etab=0.1$, $\itp=11$ with age $\tev=30$ yr is $\delta_\mathrm{L} = 0.23$.
The use of this model as initial conditions for calculations of self--excited nonlinear stellar oscillations
with time--independent inner boundary conditions leads to the limit cycle solution in the form of the first
overtone pulsator with period $\Pi=108$ d.
However, as seen in Fig.~\ref{fig:fig5}, the hydrodynamic model with age $\tev=30$ yr is still the fundamental
mode pulsator with period $\Pi\approx 260$ d.

\subsection{Pulsational mode switching}
\label{sec:pulmodsw}

\begin{figure}
	\includegraphics[width=\columnwidth]{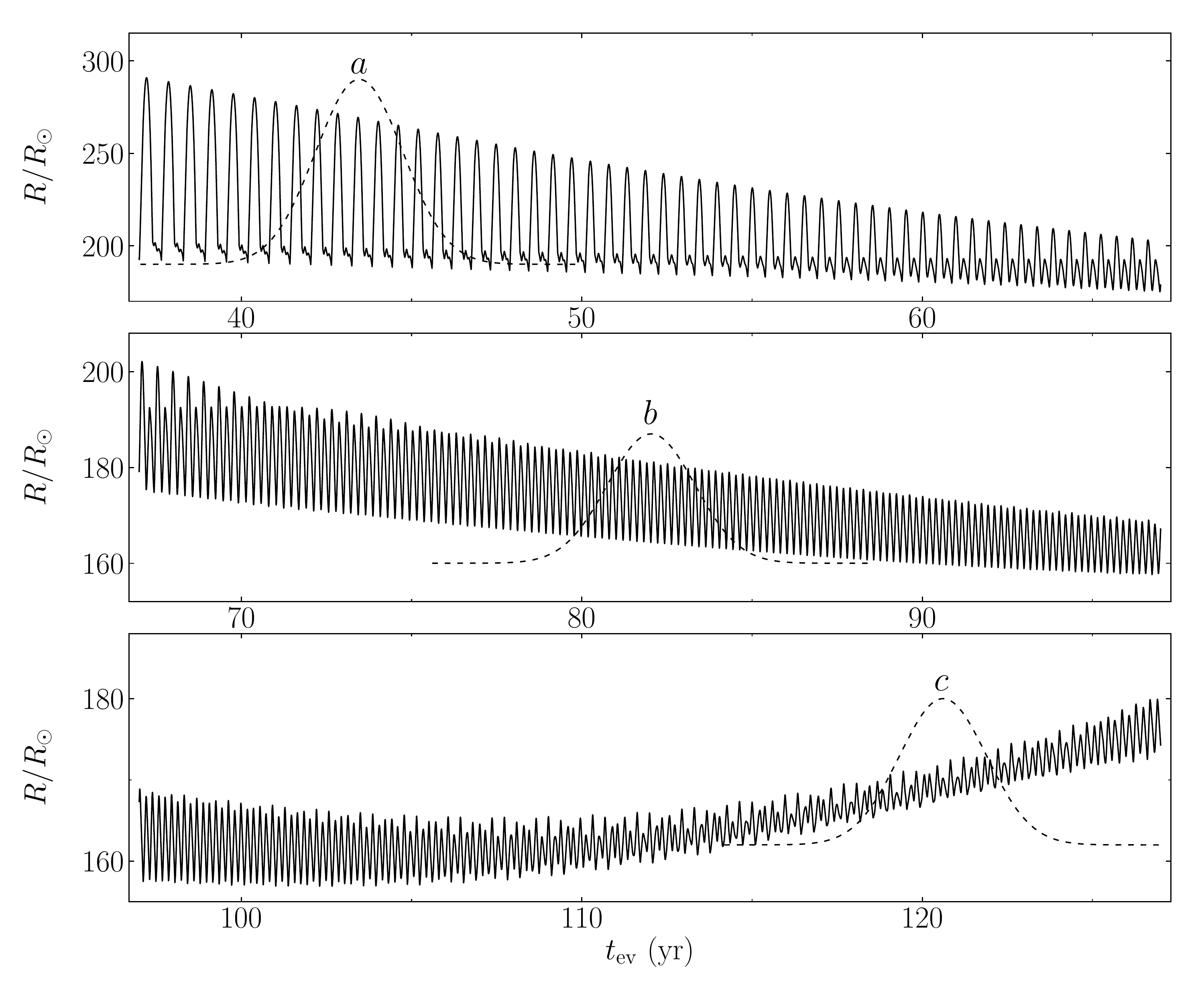}
    \caption{The radius of the outer boundary of the hydrodynamic model
    $\Mzams=1.8M_{\sun}$, $\etab = 10$, $\itp = 11$
    as a function of time during pulsational mode switching.
    Dashed lines represent the location of time intervals (gaussian windows)
    used for calculation of the short--time Fourier transform.}
    \label{fig:fig8}
\end{figure}

Solution of the equations of hydrodynamics with time--dependent inner boundary conditions
allowed us to trace mode switching in the contracting stellar envelope.
This is illustrated in Fig.~\ref{fig:fig8} where the radius of the outer boundary
of the hydrodynamic model $\Mzams=1.8M_{\sun}$, $\etab = 10$, $\itp = 11$ is shown
as a function of time $\tev$.
Mode switching from the fundamental mode to the first overtone occurs in the time interval
$30~\textrm{yr}\loa\tev\loa 70~\textrm{yr}$.
The period ratio of the first overtone and the fundamental mode is $\Pi_1/\Pi_0\approx 0.50$
(see Table~\ref{tab:table}) so that mode switching is seen as the gradually increasing bump
between two adjacent maxima of the radius.
All hydrodynamic models show the similar behaviour and the only difference is the age
at the onset of mode switching.

Mode switching from the fundamental mode to the first overtone is due to temperature growth
in the pulsating envelope because it is accompanied by outward movement of the lower boundary
of the hydrogen ionization zone in the first three decades after the peak helium--burning luminosity.
Subsequent temperature decrease in the stellar envelope ($\tev\goa 30$ yr) leads to inward movement
of the lower boundary of the hydrogen ionization zone and to mode switching from the first overtone
to the fundamental mode for $\tev\goa 100$ yr.

\begin{figure}
	\includegraphics[width=\columnwidth]{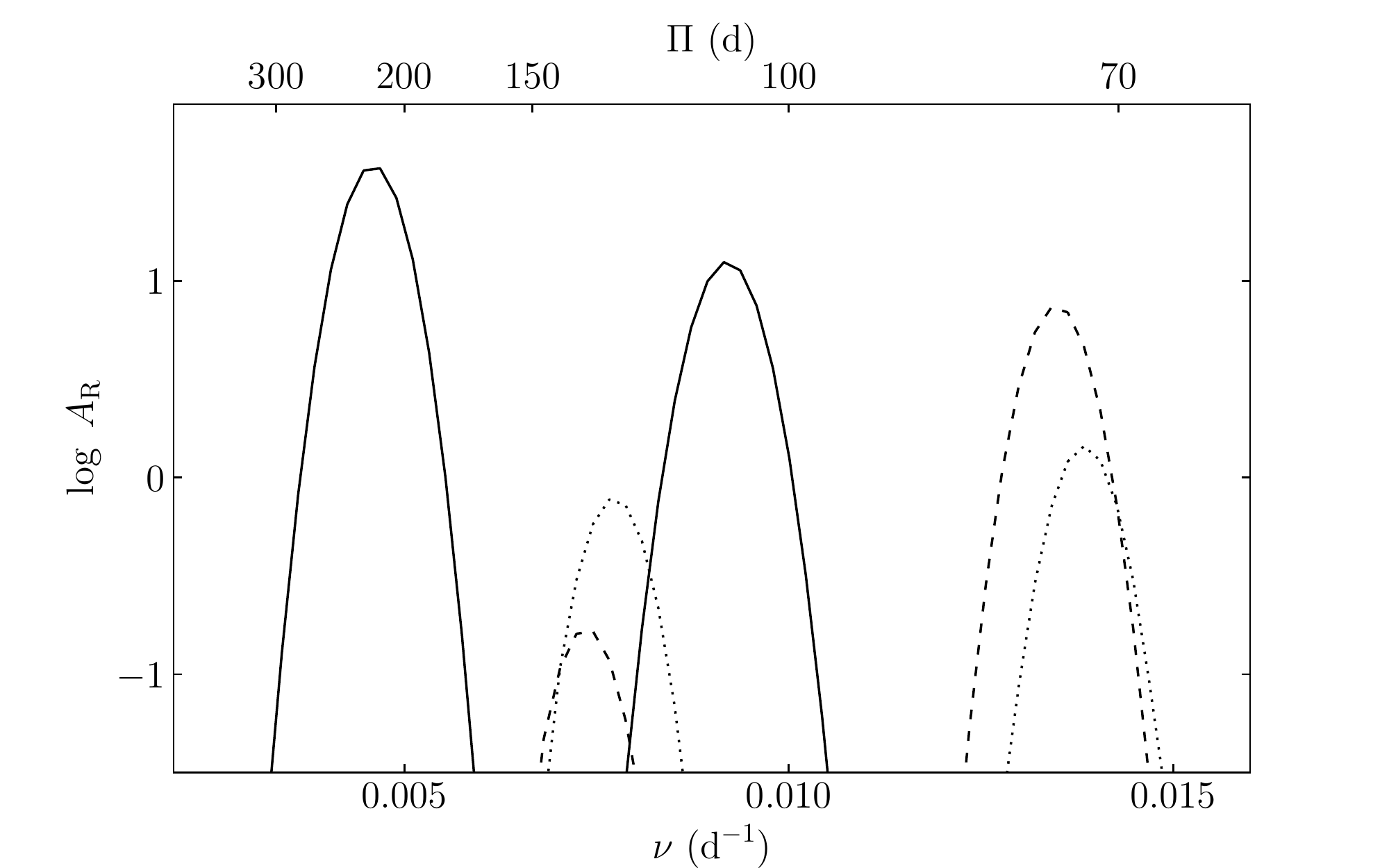}
    \caption{
    The amplitude spectra $A_\mathrm{R}$ for time intervals labelled in Fig.~\ref{fig:fig8}
    as $a$ (solid lines), $b$ (dashed lines) and $c$ (dotted lines).
    The low--frequency and the high--frequency maxima of the spectra correspond to the fundamental
    mode and the first overtone, respectively.}
    \label{fig:fig9}
\end{figure}

Mode switching in the hydrodynamic model of T UMi can be also illustrated by results of the short--time
Fourier transform of the temporal dependence of the stellar radius presented in Fig.~\ref{fig:fig8}.
To this end the whole time interval of the hydrodynamic model
$\Mzams=1.8M_{\sun}$, $\etab = 10$, $\itp = 11$ ($-70~\textrm{yr}\le \tev\le 127$~\textrm{yr})
was divided into 16 time intervals and the amplitude spectrum $A_\mathrm{R}$ was computed for the product
of $R(\tev)$ and the gaussian window function $\sigma=5$ for each interval.
The amplitude spectra of the stellar radius $A_\mathrm{R}$ are given in Fig.~\ref{fig:fig9}
for three intervals shown in Fig.~\ref{fig:fig8} by the dashed lines.
The case '$a$' (solid lines) corresponds to the early stage of mode switching when the amplitude
of the first overtone is nearly one third of that of the fundamental mode whereas in the case '$b$'
(dashed lines) the star became the first overtone pulsator since the spectral amplitude of the
first overtone is about forty times greater in comparison with that of the fundamental mode.
The amplitude spectrum shown in Fig.~\ref{fig:fig9} by dotted lines (the case '$c$') allows us
to conclude that increasing scatter of the curve $h_1$ in the upper panel of Fig.~\ref{fig:fig5}
corresponds to the beginning stage of a modal switch from the fundamental mode to the first overtone pulsation.

\begin{figure}
	\includegraphics[width=\columnwidth]{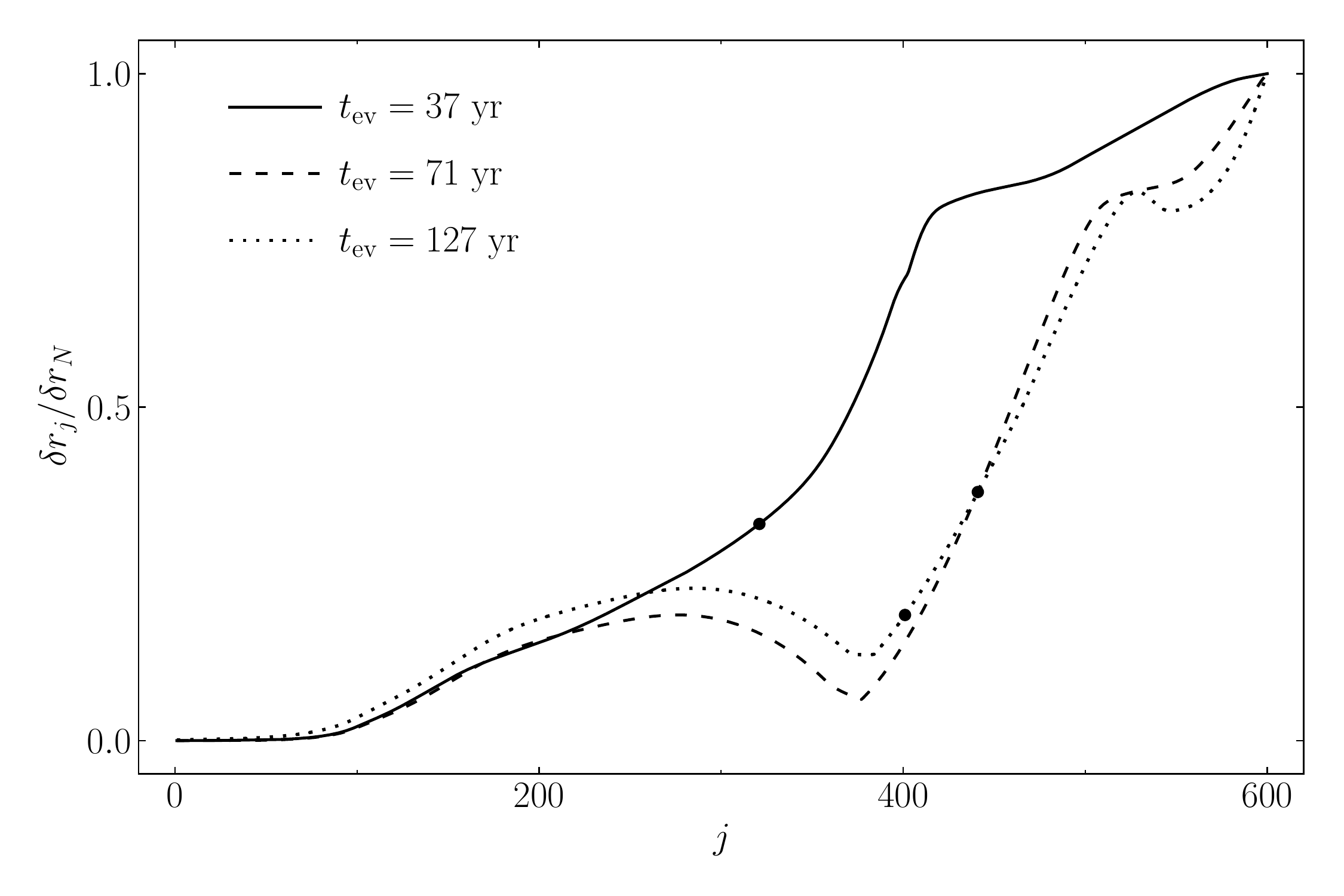}
    \caption{The normalized amplitude of the radial displacement of the $j$--th mass zone
    in the hydrodynamic model $\Mzams=1.8M_{\sun}$, $\etab=0.1$, $\itp=11$ for
    $\tev=37$~yr (solid line), $\tev=71$~yr (dashed line) and $\tev=115$ yr (dotted line).
    The mass zone $j=0$ corresponds to the inner boundary of the hydrodynamic model.
    Filled circles roughly indicate the inner boundary of the
    hydrogen ionization zone.}
    \label{fig:fig10}
\end{figure}

Fig.~\ref{fig:fig10} shows the plots of the radial displacement of Lagrangean mass zones
in the hydrodynamic model $\Mzams=1.8M_{\sun}$, $\etab=0.1$, $\itp=11$ for three
values of the evolution time $\tev$.
For the sake of graphical representation the plots represent the normalized amplitudes
$\delta r_j/\delta r_N$ because the surface amplitudes $\delta r_N$ significantly differ from
each other.
The plot for $\tev=37$~yr corresponds to regular oscillations in the fundamental mode,
whereas at the evolution time $\tev=71$~yr the star oscillates in the first overtone.
The node of the first overtone locates at the mass zone $j_\mathrm{n}\approx 380$ with mean radius
$\langle r\rangle\approx 0.80\langle r_N\rangle$, whereas the inner boundary of
the hydrogen ionization zone locates in mass zones $j > j_\mathrm{n}$.
The non--zero amplitude of the radial displacement at the overtone node is due to the fact
that pulsation motions are
not perfectly periodic due to the presence of more than one mode.
The plot for $\tev=126$~yr corresponds to semi--regular oscillations and pulsational mode
switching from the first overtone to the fundamental mode.

\section{Conclusions}

Calculations of non--linear stellar pulsations with time--dependent inner boundary conditions
implicitly take into account effects of deviation from thermal equilibrium in the envelope
of the pulsating star.
As mentioned above, before the onset of period decrease ($\tev < 0$) deviation from thermal
equilibrium can be neglected ($\delta_\mathrm{L} \loa 0.02$).
However, for $\tev > 0$ deviation from thermal equilibrium rapidly increases and becomes as
high as $\delta_\mathrm{L} \loa 0.3$ so that application of the theory of stellar pulsations
with time--independent inner boundary conditions (\ref{eq:sibc}) leads to wrong solutions.
The difference between these solutions is due to the fact that growth of the amplitude
in the model computed with time--independent inner boundary conditions is accompanied by changes
in the entropy structure of the pulsating star and relaxation of the model to the
configuration with smaller stellar radius \citep{yt1996}.

The hydrodynamic models of T UMi listed in Table~\ref{tab:table} show good agreement with
observational estimates of the period and the rate of period change.
At the same time they are characterized by a relatively wide range of stellar masses:
$1.04M_{\sun}\le M\le 1.48M_{\sun}$.
To provide a stronger constraint on the stellar mass of T UMi one has to fix the date
when the pulsation period ceases to decrease.
Additional constraints on the model parameters can be obtained from observational
estimates of the pulsation amplitude near the period minimum.

\section*{Data Availability}

The data underlying this article will be shared on reasonable request to the corresponding author.



\bibliographystyle{mnras}
\bibliography{fadeyev}


\bsp	
\label{lastpage}
\end{document}